\documentclass[12pt]{article}
\usepackage[latin9]{inputenc}
\usepackage{float}
\usepackage{mathtools}
\usepackage{amsmath}
\usepackage{amssymb}
\usepackage{cancel}
\usepackage{stmaryrd}
\usepackage{stackrel}
\usepackage{undertilde}
\usepackage{graphicx}
\usepackage{esint}



\begin{document}
\title{Toward ultrametric modeling of the epidemic spread }
\author{V.\,T.~Volov \\
 \textit{Natural Science Department, } \\
 \textit{Samara State University of Railway Transport,} \\
 \textit{Perviy Bezimyaniy pereulok 18, 443066, Samara, Russia} \\
 e-mail:\:\texttt{volovvt@mail.ru} \\
 and \\
 A.\,P.~Zubarev \\
 \textit{ Physics Department, Samara University, } \\
 \textit{ Moskovskoe shosse 34, 443123, Samara, Russia} \\
 \textit{Natural Science Department,} \\
 \textit{Samara State University of Railway Transport,} \\
 \textit{Perviy Bezimyaniy pereulok 18, 443066, Samara, Russia} \\
 e-mail:\:\texttt{apzubarev@mail.ru} }
\maketitle
\begin{abstract}
An ultrametric model of epidemic spread of infections based on the
classical SIR model is proposed. Ultrametrics on a set of individuals
is introduced based on their hierarchical clustering relative to the
average time of infection contact. The general equations of the ultrametric
SIR model are written down and their particular implementation using
the $p$-adic parametrization is presented. A numerical analysis of
the $p$-adic SIR model and a comparison of its behavior with the
classical SIR model are performed. The concept of hierarchical isolation
and the scenario of its management in order to reduce the level of
epidemic spread is considered.

Keywords: SIR model, hierarchical clustering, ultrametrics, $p$-adic
models, epidemic spread.
\end{abstract}

\section{Introduction}

Recently, there has been activity in the development of mathematical
models of the spread of infectious diseases (see, for example, \cite{AM,DH,KR,BS,WB}).
This is primarily due to both the detection of old foci of existing
viral infections, such as plague, anthrax, Ebola, and the emergence
of the new infections caused by coronaviruses such as SARS-CoV (2002),
MERS-CoV (2015), and SARS-CoV-2 (2019). Almost all mathematical models
describing the development of epidemics are based on the model proposed
in 1927 by Kermack and McKendrick in work \cite{KM}, which is now
known as the classic SIR model. This model is based on 3-step development
of the epidemic, in which healthy but susceptible individuals (S)
as a result of infection go to the infected class (I) individuals
who, in turn, move to the removed class (R), i.e. recover by acquiring
immunity, or die. Note that the SIR model was constructed by analogy
with the theory of homogeneous chemical reactions. This model has
many generalizations that include additional intermediate stages of
individuals known as SIRS, SEIR, SEIRS, and others.

The essence of the simplest SIR model is as follows. Let $S\left(t\right)$
is the number of susceptible, $I\left(t\right)$ is the number of
infected, $R\left(t\right)$ is the number of removed individuals.
The following simplifying assumptions are accepted: 1) the set of
individuals is homogeneous; 2) the number of individuals under consideration
(susceptible, infected and removed) is constant, i.e. $S\left(t\right)+I\left(t\right)+R\left(t\right)=N$;
3) the probability of transmission per unit time from an infected
individual to a susceptible individuals is constant; 3) the probability
of reinfection is zero. Given these simplifying assumptions the SIR
model equations are written as

\begin{equation}
\dot{S}=-\dfrac{\beta}{N}SI,\label{S1}
\end{equation}

\begin{equation}
\dot{I}=\dfrac{\beta}{N}SI-\gamma I,\label{I1}
\end{equation}

\begin{equation}
\dot{R}=\gamma I.\label{R1}
\end{equation}
Here $\beta$ is infection rate (or the average number of contacts
with susceptible individuals that leads to the new infected individuals
per time unit per infective), and $\gamma$ is removing rates (or
the average rate of removal of infective per unit time per infective).

The joint solution of equations (\ref{S1}) and (\ref{R1}) gives

\begin{equation}
S=S_{0}\exp\left(-\dfrac{\mathrm{\mathcal{R}_{0}}}{N}\cdot R\right),\label{S(R)}
\end{equation}
where $S\left(0\right)\equiv S_{0}$ and $\mathrm{\mathcal{R}_{0}}=\dfrac{\beta}{\gamma}$
is reproductive ratio. It follows from (\ref{S(R)}) that
\begin{equation}
1-\dfrac{R_{\infty}}{N}=\dfrac{S_{0}}{N}\exp\left(-\mathrm{\mathrm{\mathcal{R}_{0}}}\dfrac{R_{\infty}}{N}\right),\label{R}
\end{equation}
where $R_{\infty}\equiv\lim_{t\rightarrow\infty}R\left(t\right)$.
Equation (\ref{R}) with $\dfrac{S_{0}}{N}=1$ is solvable for $\dfrac{R_{\infty}}{N}$
on the segment $\left(0,\:1\right)$ only under condition $\mathrm{\mathrm{\mathcal{R}_{0}}}>1$
and there is no solutions on this segment when $\mathrm{\mathrm{\mathcal{R}_{0}}}<1$.
For this reason, condition $\mathrm{\mathrm{\mathcal{R}_{0}}}>1$
is interpreted in this model as an outbreak condition of epidemic.
Today, many generalizations of the classic SIR model have been proposed,
which use more complex scenarios for the development of the epidemic,
taking into account the latency period, the finiteness of the time
when individuals are in the stage of immunity, the impact of vaccination,
etc. The classical SIR model, as well as most of its generalizations,
are essentially homogeneous models, i.e. they assume that the intensity
of infection does not depend on any relationship between susceptible
and infected individuals. Nevertheless, even in such homogeneous models,
adequate estimates of the scenario for the development of real epidemics
are possible. Recently, a number of heterogeneous generalizations
of the SIR models based on deterministic or random networks have also
been proposed (see, for example, \cite{LM,MPV,YWRBSWZ,Volz}), taking
into account the heterogeneous nature of the populations in which
the epidemic is spreading. As a rule, this heterogeneity is associated
with the difference in the characteristics of individuals, or with
the heterogeneity of population density.

Meanwhile, the most important characteristic that plays a crucial
role in the speed of spread of the epidemic is the distribution of
the time duration of infectious contact between pairs of individuals.
This value can be a characteristic of population heterogeneity, and
it allows us to introduce a distance function on a set of individuals
(metric) in terms of which it is possible to generalize the classical
SIR model.

In this paper, we propose an inhomogeneous generalization of the classical
SIR model, which is based on the hierarchical clustering of the population
according to the degree of potential infectious influence of individuals
on each other. We describe the procedure for such hierarchical clustering
of a set of individuals and show that such clustering involves the
introduction of a distance function on the set of individuals, which
is ultrametrics. This distance function is not directly related to
the spatial distance between individuals, it is determined only by
the potential for transmission of infection from one individual to
another per unit time. Using the ultrametric distance on a set of
individuals, we generalize the equations of the classical SIR model
and arrive at a model that we call the basic ultrametric SIR model.
As known, a convenient tool for parameterizing ultrametric spaces
is the field of $p$-adic numbers. We consider the ultrametric set
of individuals that can be maped to the boundary of a finite hierarchical
graph with a constant number of branches. Such a set can be parameterized
by the set of a $p$-adic balls of unit radius contained in a $p$-adic
ball of radius greater than 1. We call the corresponding ultrametric
SIR model the $p$-adic SIR model. We present a numerical analysis
of the $p$-adic SIR model and compare its behavior with the classical
SIR model. We also introduce the concept of hierarchical isolation
index and consider the simplest scenario for managing this parameter
in order to reduce the spread of the epidemic.

\section{Hierarchical clustering of human population and ultrametric}

Let $U=\left\{ x\right\} $ be a set of $N$ individuals. Infectious
contact $A$ of two individuals $x$ and $y$ we will call any continuous
time interval $A=\left(t_{1},\:t_{2}\right)$ which satisfies the
condition: during interval $A$ one of the individuals (for example,
$x$) can be infected by another individual $y$, provided that individual
$y$ is infected throughout the entire interval $A$, and the probability
of infection per unit time $f\left(t\right)$ is non zero on interval
on $A$ almost everywhere.

Note that the concept of an infectious contact is a purely model.
In almost every specific case, the value of the infective contact
is extremely difficult to calculate accurately, but it is always possible
to approximate it. However, in any real assessment of the duration
of an infectious contact, it must be understood that this concept
does not necessarily mean physical contact between two individuals.
The probability of infection of one individual by another individual
may occur indirectly, i.e. through a physical contact with surrounding
bodies (air, objects) that became sources of infection after a physical
contact with an infected individual. However, an infectious contact
does not imply the possibility of infection of a susceptible individual
by an already infected individual through any other individuals. Intuitively
speaking, the term ``infectious contact'' is a continuous period
of time during which a particular individual has the potential for
infection (no matter how small it is) at any time, and this possibility
of infection is caused by a particular other individual.

Let $T_{0}$ be the value of a certain time interval during which
we observe the behavior of individuals. Then for individual $x$,
there is a finite number of infectious contacts $A_{1},\:A_{2},\:\ldots,A_{k}$,
$k\geq0$ with individual $y$ during this interval. Denote by $T\left(x,y\right)$
the sum of the interval lengths $A_{1},\:A_{2},\:\ldots,A_{k}$. Obviously
always $T\left(x,y\right)\leq T_{0}$. Since the behavior of individuals
is random, $T\left(x,y\right)$ is a random function which depends
on the specific implementation of an ensemble of populations of individuals.
Expected value $\tau\left(x,y\right)=\mathbb{E}\left[T\left(x,y\right)\right]$
is the average time of an infectious contact between two individuals
during interval $T_{0}$. It is obvious that the function $\tau\left(x,y\right)$
is symmetric.

Function $\tau\left(x,y\right)$ plays the main role in the hierarchical
clustering of individuals, which we describe below. Consider an infinite
decreasing sequence of numbers of real positive numbers $t_{1},\:t_{2},\:t_{3},\:\ldots$,
such that $\dfrac{t_{i}}{t_{i+1}}\ll1$. Let's divide all the set
of individuals $U$ into groups $B_{1}^{(1)}$, $B_{1}^{(2)}$, ...,
$B_{1}^{(n_{1})}$, $\bigcup_{i=1}^{n_{1}}B_{1}^{(i)}=U$ as follows.
We will require that for any $i$ and for any individual $x\in B_{1}^{(i)}$
there exists another individual $y\in B_{1}^{(i)}$ from the same
group for which inequality $\tau\left(x,y\right)\geq t_{1}$ holds.
Obviously, this partition is the only one. It then follows that if
$i\neq j$, then $\forall\:x\in B_{1}^{(i)}$ and $\forall\:y\in B_{1}^{(j)}$
then we have $\tau\left(x,y\right)<t_{1}$. Groups $B_{1}^{(i)}$
will be called the first level clusters.

First level clusters $\left\{ B_{1}^{(1)},B_{1}^{(2)},\ldots,B_{1}^{(n_{1})}\right\} $
can be combined into larger sets of individuals -- the second level
clusters, which we will denote by $B_{2}^{(1)}$, $B_{2}^{(2)}$,
..., $B_{2}^{(n_{2})}$, $\bigcup_{i=1}^{n_{2}}B_{2}^{(i)}=U$, $n_{2}<n_{1}$.
In this case, we require that for any individual $x\in B_{2}^{(i)}$
there exists another individual $y\in B_{2}^{(i)}$ from the same
cluster for which inequality $\tau\left(x,y\right)\geq t_{2}$ holds.
It is obvious that if $i\neq j$, then $\forall\:x\in B_{2}^{(i)}$
and $\forall\:y\in B_{2}^{(j)}$ then $\tau\left(x,y\right)<t_{2}$.
Also for any $i$ if $x\in B_{2}^{(i)}$ and $y\in B_{2}^{(i)}$ we
have $\tau\left(x,y\right)\geq t_{1}$ if $x$ and $y$ belong to
the same first level cluster and $t_{2}\leq\tau\left(x,y\right)<t_{1}$
if $x$ and $y$ belong to different first level clusters. Similarly
the clusters of the second level can be combined into third level
clusters $B_{3}^{(1)}$, $B_{3}^{(2)}$, ..., $B_{3}^{(n_{3})}$,
$\bigcup_{i=1}^{n_{3}}B_{3}^{(i)}=U$, $n_{3}<n_{2}$. For any individual
$x\in B_{3}^{(i)}$, there exists another individual $y\in B_{3}^{(i)}$
from the same cluster for which inequality $\tau\left(x,y\right)\geq t_{3}$
holds. The third level clusters can be combined into fourth level
clusters $B_{4}^{(1)}$, $B_{4}^{(2)}$, ..., $B_{4}^{(n_{4})}$,
$\bigcup_{i=1}^{n_{4}}B_{4}^{(i)}=U$, $n_{4}<n_{3}$ , and so on.
This procedure of hierarchical clustering of individuals must eventually
be interrupted, because at some $n$-th step we will get a single
cluster of $n$-th level $B_{n}^{(1)}$ that will coincide with the
set of individuals $U$. For any individual $x\in B_{n}^{(i)}\equiv U$,
there exists an individual $y\in B_{n}^{(i)}\equiv U$ for whisch
$\tau\left(x,y\right)\geq t_{n}$ holds.

So, the hierarchical procedure for clustering individuals described
above allows us to introduce on set $U$ the structure of hierarchically
nested clusters $B_{i}^{(j)}$ . By construction, any two clusters
of arbitrary levels either do not intersect, or one is contained in
the other. The resulting hierarchical structure of clusters at various
levels can be described by a hierarchical graph. An example of such
a graph is shown in Figure 1.

\begin{figure}[hbt!]
\centering{}\includegraphics[scale=0.5]{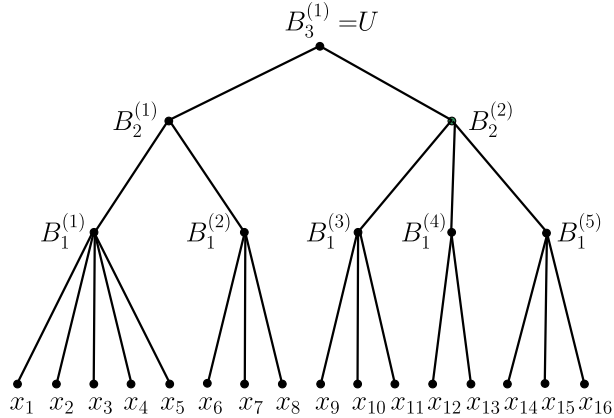}\caption{An example of a hierarchical graph corresponding to a 3-level hierarchical
clustering of a set of 16 individuals.}
\end{figure}

Hierarchical clustering of the set of individuals $U$ allows us to
introduce the ultrametrics $d\left(x,y\right)$ on this set. We define
function $d\left(x,y\right)$ in the following way. If individuals
$x$ and $y$ belong to the same level $i$ cluster, but belong to
different level $i-1$ subclusters, then we put $d\left(x,y\right)=\dfrac{T_{0}}{t_{i}}$.
If $x=y$, we will put $d\left(x,y\right)=0$.

\textbf{Proposition.} \emph{Function $d\left(x,y\right)$ is ultrametric,
i.e. it satisfies the strong triangle inequality
\begin{equation}
\forall\:x,y,z\:d\left(x,y\right)\leq\max\left\{ d\left(y,z\right),d\left(x,z\right)\right\} .\label{tr_ne}
\end{equation}
}

\textbf{Proof. } To prove that function $d\left(x,y\right)$ is ultrametric,
it is sufficient to prove that any ``triangle'' $\left(x,y,z\right)$,
where $x\in U$, $y\in U$, $z\in U$ is isosceles, and the two largest
``sides'' are equal. Let $d\left(x,y\right)=\max\left\{ d\left(x,y\right),d\left(y,z\right),d\left(x,z\right)\right\} $.
Then by definition $d\left(x,y\right)$, individuals $x$ and $y$
belong to some level $i$ cluster $B^{(i)}$, but they belong to different
subclusters $B_{1}^{(i-1)}$ and $B_{2}^{(i-1)}$ level $i-1$. Next,
we have $d\left(y,z\right)\leq d\left(x,y\right)$, $d\left(x,z\right)\leq d\left(x,y\right)$.
Suppose that inequalities $d\left(y,z\right)<d\left(x,y\right)$ and
$d\left(x,z\right)<d\left(x,y\right)$ hold. Then each of the pairs
$\left(y,z\right)$ and $\left(x,z\right)$ belongs to some cluster
of a level smaller than $i$. Let $\left(y,z\right)\in B_{1}^{(j)}$,
$\left(y,z\right)\in B_{1}^{(k)}$, where $j,k<i$. Both clusters
$B_{1}^{(j)}$ and $B_{1}^{(j)}$ contain the same element $y$ .
Therefore, one of them is contained in the other. Then the larger
cluster must contain all three individuals $x,y,z$. But this means
that $x$ and $y$ are contained in a cluster of a level smaller than
$i$. Therefore, inequalities $d\left(y,z\right)<d\left(x,y\right)$
and $d\left(x,z\right)<d\left(x,y\right)$ cannot simultaneously be
satisfied. Hence either $d\left(y,z\right)=d\left(x,y\right)$, or
$d\left(x,z\right)=d\left(x,y\right)$. The proposition is proven.

\section{Ultrametric formulation of the SIR model}

In its meaning, ultrametrics $d\left(x,y\right)$ on a set of individuals
is a value equal to the ratio of the population observation time to
the minimum boundary value of the average times of infectious contact
of any two individuals belonging to the same minimal cluster as individuals
$x$ and $y$. We make the model assumption that the probability of
infection of a susceptible individual $x$ by an infected individual
$y$ per unit time is proportional to value $\dfrac{1}{d\left(x,y\right)}$.
Naturally, this assumption is rather a rough approximation, since
it assumes that during the infectious contact of individuals, the
probability of infection per unit time $f\left(t\right)$ of a susceptible
individual infected is constant. Naturally, real function $f\left(t\right)$
is not constant and depends on many factors (contact space, pathogen
concentration, individual susceptibility, etc.). However, this approximation
allows us to simplify the model significantly, while preserving its
main quality property -- the hierarchical nature of the development
of the epidemic.

Let $P_{I}\left(x,t\right)$, $P_{R}\left(x,t\right)$, $P_{S}\left(x,t\right)$
be the probability that at time $t$, individual $x$ is correspondingly
susceptible, infected, or removed. Then we can write the following
ultrametric generalization of the equations (\ref{S1}) -- (\ref{R1})

\begin{equation}
\dot{P_{S}}\left(x,t\right)=-\widetilde{\beta}P_{S}(x,t)\sum_{y,y\neq x}\dfrac{1}{d\left(x,y\right)}P_{I}(y,t),\label{SU}
\end{equation}

\begin{equation}
\dot{P_{I}}\left(x,t\right)=\widetilde{\beta}P_{S}(x,t)\sum_{y,y\neq x}\dfrac{1}{d\left(x,y\right)}P_{I}(y,t)-\gamma P_{I}\left(x,t\right),\label{IU}
\end{equation}

\begin{equation}
\dot{P_{R}}\left(x,t\right)=\gamma P_{I}\left(x,t\right).\label{RU}
\end{equation}
Function
\begin{equation}
W\left(x,y\right)=\dfrac{\widetilde{\beta}}{d\left(x,y\right)}\label{rate}
\end{equation}
makes sense of the probability of infection by an infected individual
$y$ of a susceptible individual $x$ per unit time. For large $N$,
the total number of susceptible, infected, and removed individuals
is

\[
S(t)=\sum_{x}P_{S}(x,t),\;I(t)=\sum_{x}P_{I}(x,t),\;R(t)=\sum_{x}P_{R}(x,t).
\]

In order to have a connection with the equations (\ref{S1}) -- (\ref{R1}),
it is convenient to put

\begin{equation}
\widetilde{\beta}=\dfrac{\beta}{N}.\label{tilde_beta}
\end{equation}
Then in the case of the triviality of ultrametrics

\[
d\left(x,y\right)=\left\{ \begin{array}{c}
1,\:x\neq y,\\
0,\:x=y,
\end{array}\right.
\]
and the independence of the functions $P_{I}\left(x,t\right)$, $P_{R}\left(x,t\right)$,
$P_{S}\left(x,t\right)$ from variable $x$, taking into account $S(t)=NP_{S}(x,t)$,
$S(t)=NP_{S}(x,t)$, $R(t)=NP_{R}(x,t)$, we get the system of equations

\[
\dot{S}=-\dfrac{\beta\left(N-1\right)}{N^{2}}SI,
\]

\[
\dot{I}=\dfrac{\beta\left(N-1\right)}{N^{2}}SI-\gamma I,
\]

\[
\dot{R}=\gamma I,
\]
which for large $N$ in approximation $\dfrac{N-1}{N}\approx1$ coincides
with the system (\ref{S1}) -- (\ref{R1}). We will call the system
(\ref{SU}) -- (\ref{RU}) the equations of the basic ultrametric
SIR model.

\section{$p$-adic parametrization of the ultrametric SIR model}

Recall the definition of $p$-adic number. Let $\mathbb{Q}$ be a
field of rational numbers and let $p$ be a fixed prime number. Any
rational number $x\ne0$ is uniquely represented as

\begin{equation}
x=\pm p^{\gamma}\frac{a}{b},\label{x_p}
\end{equation}
where $\gamma$ is an integer, and $a$, $b$ are natural numbers
that are not divisible by $p$ and have no common multipliers. The
$p$-Adic norm $\left|x\right|_{p}$ of number \textit{$x\in\mathbb{Q}$}
is defined by the equalities $\left|x\right|_{p}=p^{-\gamma}$, $\left|0\right|_{p}=0$.
The field of $p$-adic numbers $\mathbb{Q}_{p}$ is defined as a completion
of the field of rational numbers $\mathbb{Q}$ by $p$-adic norm $\left|x\right|_{p}$.
The norm on $\mathbb{Q}_{p}$ induces the metric $d(x,y)=\left|x-y\right|_{p}$
which is ultrametric, i.e. satisfies the strong triangle inequality
(\ref{tr_ne}). We will denote: $B_{i}(a)=\{x\in\mathbb{Q}_{p}:\:|x-a|_{p}\leq p^{i}\}$
-- a ball of radius $p^{i}$ centered at point $a$, $S_{i}(a)=\{x\in\mathbb{Q}_{p}:\:|x-a|_{p}=p^{i}\}$
-- a sphere of radius $p^{i}$ centered at point $a$, $B_{i}\equiv B_{i}(0)$,
$S_{i}\equiv S_{i}(a)$, $\mathbb{Z}_{p}\equiv B_{0}$. On $\mathbb{Q}_{p}$
there exists a~unique (up to a~factor) Haar measure $d_{p}x$ which
is invariant with respect to translations $d_{p}\left(x+a\right)=d_{p}x$.
We assume that $d_{p}x$ is a~full measure; that is,
\begin{equation}
\intop_{\mathbb{Z}_{p}}d_{p}x=1.\label{norm}
\end{equation}
Under this hypothesis the measure $d_{p}x$ is unique. For more information
about $p$-adic numbers, the $p$-adic analysis and its applications,
see \cite{ALL,VVZ,Sh}.

For our purposes, we can assume that number $p$ is a natural number
$p=m>2$. In this case $\mathbb{Q}_{p}$ is a ring of $m$-adic numbers
$\mathbb{Q}_{m}$ with the pseudo-norm $\left|x\right|_{m}$, which
also induces on $\mathbb{Q}_{m}$ the ultrametrics $d(x,y)=\left|x-y\right|_{p}$
\cite{DZ}. So let $p\geq2$ be a natural number. Further, let each
level $i$ cluster contains exactly $p$ level $i-1$ clusters, the
number of levels is $n$, and the total number of individuals is $N=p^{n}$.
In this case, the set of individuals $U$ can be parameterized by
the factor set $B_{n}/\mathbb{Z}_{p}$ and we assume that $U=B_{n}/\mathbb{Z}_{p}$.
Alternatively, we can describe the set of individuals by $B_{n}\subset\mathbb{Q}_{p}$,
but assume that each individual is described by a ball of unit radius.

Let $D\left(\lambda\right)$ be an arbitrary non-negative non-decreasing
function defined on $\mathbb{R}_{+}$ and satisfy the condition $D\left(0\right)=0$.
Then function $D\left(\left|x-y\right|_{p}\right)$ is also an ultrametrics
on $B_{n}$ and we can write the equations of the basic ultrametric
SIR model (\ref{SU}) -- (\ref{RU}) in the form:

\begin{equation}
\dot{P_{S}}\left(x,t\right)=-\widetilde{\beta}P_{S}(x,t)\intop_{B_{r}}d_{p}y\dfrac{P_{I}(y,t)-\Omega(\left|y-x\right|_{p})P_{I}(x,t)}{D\left(\left|x-y\right|_{p}\right)},\label{pSU}
\end{equation}

\begin{equation}
\dot{P_{I}}\left(x,t\right)=\widetilde{\beta}P_{S}(x,t)\intop_{B_{r}}d_{p}y\dfrac{P_{I}(y,t)-\Omega(\left|y-x\right|_{p})P_{I}(x,t)}{D\left(\left|x-y\right|_{p}\right)},-\gamma P_{I}\left(x,t\right),\label{pIU}
\end{equation}

\begin{equation}
\dot{P_{R}}\left(x,t\right)=\gamma P_{I}\left(x,t\right),\label{pRU}
\end{equation}
where

\[
\Omega(\left|x\right|_{p}p^{i})=\left\{ \begin{array}{l}
1,\mathrm{\;}\left|x\right|_{p}p^{i}\leq1,\\
0,\mathrm{\;}\left|x\right|_{p}p^{i}>1.
\end{array}\right.
\]
As function $D\left(\left|x-y\right|_{p}\right)$ in equations (\ref{pSU})
-- (\ref{pRU}), we choose

\begin{equation}
D\left(\left|x-y\right|_{p}\right)=\left|x-y\right|^{\alpha}.\label{D}
\end{equation}
To preserve the interpretation, we will assume that functions $P_{S}(x,t)$,
$P_{I}(x,t)$, $P_{R}(x,t)$ lie in the class $W_{0}\cap L^{1}\left(B_{r},d_{p}x\right)\cap C^{1}\left(\mathbb{R}_{+}\right)$.
Here $W_{0}$ is the class of functions that are constant on any ball
of unit radius, $L^{1}\left(B_{r},d_{p}x\right)$ is the class of
functions that are integrable on $B_{r}$ and $C^{1}\left(\mathbb{R}_{+}\right)$
is the class of functions that are differentiable with respect to
$t$.

As each individual is described by a ball of unit radius. In this
case, the possible non-zero values of ultrametrics (\ref{D}) are
$p^{i\alpha}$, $i=1,2,\ldots n$. This means that the ratio of boundaries
of an average infection contact times of individuals belonging to
different maximum subclusters of level $i$ and $i+1$ clusters is

\begin{equation}
r=\dfrac{t_{i}}{t_{i+1}}=p^{\alpha}.\label{r}
\end{equation}
In the $p$-adic model under consideration, we will call value $r$
the hierarchical isolation index.

We will assume that coefficient $\widetilde{\beta}$ depends on $\alpha$.
We fix this dependence by requiring that the average value of $\overline{W}$
of function (\ref{rate}) over all pairs of individuals, coincides
with $\dfrac{\beta}{N}$, where $\beta$ is an infection rate of classic
SIR model. In this case we have

\[
\overline{W}=\dfrac{2}{p^{n}\left(p^{v}-1\right)}\intop_{B_{r}}d_{p}x\intop_{B_{r}}d_{p}y\dfrac{1-\Omega\left(\left|x-y\right|\right)}{\left|x-y\right|^{\alpha}}=\widetilde{\beta}\dfrac{\left(p-1\right)\left(1-p^{\left(1-\alpha\right)n}\right)}{\left(p^{n}-1\right)\left(p^{\alpha}-p\right)}.
\]
Imposing the requirement $\overline{W}=\dfrac{\beta}{N}$ we get

\[
\widetilde{\beta}\left(\alpha\right)=\beta\dfrac{\left(1-p^{-n}\right)\left(p^{\alpha}-p\right)}{\left(p-1\right)\left(1-p^{\left(1-\alpha\right)n}\right)}
\]
and $\widetilde{\beta}\left(0\right)=\beta p^{-n}$. Thus, the $p$-adic
SIR model is characterized by the following parameters: $p$ is the
number of maximum subclusters for each cluster; $n$ is the number
of levels for hierarchical clustering; $\beta$ is an infection rate
of classical SIR model; $\gamma$ is an removing rate; $\alpha=\dfrac{\log r}{\log p}$,
where $r=\dfrac{t_{i}}{t_{i+1}}$ is the ratio of the boundaries of
the average times of an infectious contact of individuals belonging
to different maximum subclusters of clusters of neighboring levels.

\section{Numerical analysis of the $p$-adic SIR model and management of hierarchical
isolation}

Let there be a single infected individual at the initial time $t=0$.
This means the following choice of initial conditions of the Cauchy
problem for equations (\ref{pSU}) -- (\ref{pRU}):
\[
P_{S}(x,0)=1-e_{0}\left(x\right),
\]
\[
P_{I}(x,0)=e_{0}\left(x\right),
\]
\[
P_{R}\left(x,0\right)=0,
\]
where
\[
e_{0}\left(x\right)=\left\{ \begin{array}{c}
1,\:x=0,\\
0,\:x\neq0,
\end{array}\right.
\]
From the structure of equations (\ref{pSU}) -- (\ref{pRU}) and
the type of function (\ref{D}), it follows that functions $P_{S}(x,t)$,
$P_{I}(x,t)$, $P_{R}(x,t)$ will be constant on subsets $S_{i}=\left\{ x\in B_{n}/\mathbb{Z}_{p}:\:\left|x\right|_{p}=p^{i}\right\} $
for $i=1,2,\ldots,n$. We denote by $e_{i}\left(x\right)$ a characteristic
function of a subset of $S_{i}$. Then functions $P_{S}(x,t)$, $P_{I}(x,t)$,
$P_{R}(x,t)$ can be decomposed by the basis of functions $\left\{ e_{0}\left(x\right),e_{1}\left(x\right),\ldots,e_{n}\left(x\right)\right\} $:
\[
P_{S}(x,t)=\sum_{i=0}^{n}a_{i}\left(t\right)e_{i}\left(x\right),
\]

\[
P_{I}(x,t)=\sum_{i=0}^{n}b_{i}\left(t\right)e_{i}\left(x\right),
\]
\[
P_{R}(x,t)=\sum_{i=0}^{n}c_{i}\left(t\right)e_{i}\left(x\right).
\]
Substituting this expansion into equations (\ref{pSU}) -- (\ref{pRU})
we get the following system

\begin{equation}
\dot{a_{i}}\left(t\right)=-\widetilde{\beta}a_{i}\left(t\right)\sum_{j=0}^{n}\dfrac{p^{j}\left(1-p^{-1}+p^{-1}\delta_{j,0}\right)-\delta_{j,i}}{\left(\max\left\{ p^{i},p^{j}\right\} \right)^{\alpha}}b_{j}\left(t\right),\label{a}
\end{equation}

\begin{equation}
\dot{b_{i}}\left(t\right)=\widetilde{\beta}p^{-n}a_{i}\left(t\right)\sum_{j=0}^{n}\dfrac{p^{j}\left(1-p^{-1}+p^{-1}\delta_{j,0}\right)-\delta_{j,i}}{\left(\max\left\{ p^{i},p^{j}\right\} \right)^{\alpha}}b_{j}\left(t\right)-\gamma b_{i}\left(t\right),\label{b}
\end{equation}

\begin{equation}
\dot{c_{i}}\left(t\right)=\gamma b_{i}\left(t\right).\label{c}
\end{equation}
The total number of susceptible, infected, and removed individuals
is

\begin{equation}
S\left(t\right)=\intop_{B_{r}}d_{p}xP_{S}(x,t)=a_{0}\left(t\right)+\left(1-p^{-1}\right)\sum_{i=1}^{n}p^{i}a_{i}\left(t\right),\label{S(t)}
\end{equation}

\begin{equation}
I\left(t\right)=\intop_{B_{r}}d_{p}xP_{I}(x,t)=b_{0}\left(t\right)+\left(1-p^{-1}\right)\sum_{i=1}^{n}p^{i}b_{i}\left(t\right),\label{I(t)}
\end{equation}

\begin{equation}
R\left(t\right)=\intop_{B_{r}}d_{p}xP_{R}(x,t)=c_{0}\left(t\right)+\left(1-p^{-1}\right)\sum_{i=1}^{n}p^{i}c_{i}\left(t\right).\label{R(t)}
\end{equation}

Below we investigate numerical solutions of equation (\ref{a}) --
(\ref{c}) by Runge-Kutta-Fehlberg 45 method. Figure 2 shows the dependence
of the cumulative number of infected individuals $C=I+R$ on time
in the classical and $p$-adic SIR models.
\begin{figure}[hbt!]
\centering{}\includegraphics[scale=0.4]{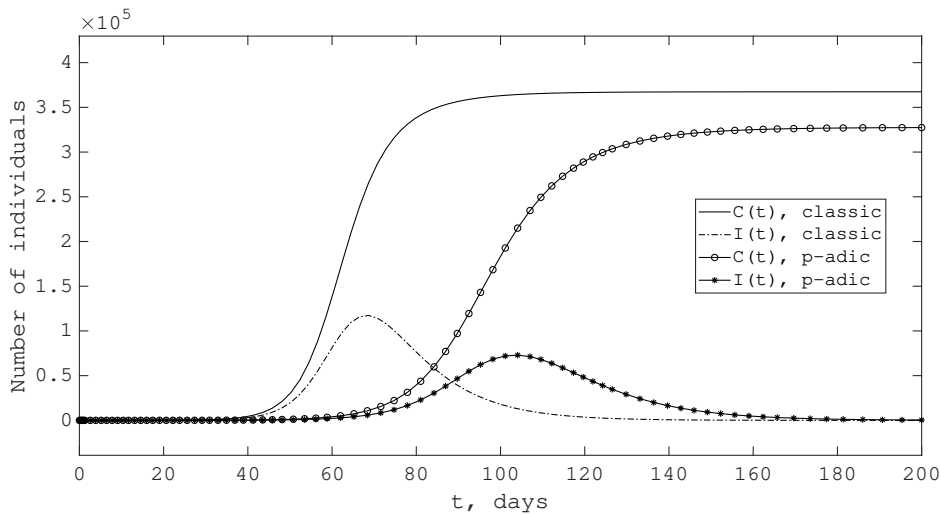}\caption{Cumulative $C\left(t\right)$ and active $I\left(t\right)$ numbers
of infected individuals in the classical and $p$-adic SIR models
for paramiters $\alpha=0.5$, $p=5$, $n=8$, $\beta=0.3$, $\gamma=0.1$. }
\end{figure}

In doing so, we have chosen $p=5$ (the number of individuals in the
minimal cluster), $n=8$ (the number of clusters), and $\alpha=0.5$.
The value of an infection rate for the classical SIR model is chosen
as $\beta=0.3$. We also choose the average time of the disease course
is equal 10 days, which corresponds to value $\gamma=0.1$. Accordingly,
the reproductive ratio for the classical SIR model is $\mathrm{\mathrm{\mathcal{R}_{0}}}=3$
(for comparison, the reproductive ratio for flu is $\mathrm{\mathrm{\mathcal{R}_{0}}}=1.3\div2.8$,
for COVID-19 $\mathrm{\mathrm{\mathcal{R}_{0}}}=2.2\div5.7$ \cite{SLXRHK}).
For these parameters, we have $r=$2.24 and the ratio of the maximum
and minimum average time of an infection contact between individuals
is $\dfrac{t_{1}}{t_{8}}=625$. The initial conditions of the Cauchy
problem of equations (\ref{a}) - (\ref{c}) are chosen as $a_{i}\left(0\right)=1-\delta_{i,0}$,
$b_{i}\left(0\right)=\delta_{i,0}$. This corresponds to one infected
individual at the initial time.

Figure 3 and Figure 4 show similar dependencies for values $\alpha=0.8$
and $\alpha=0.9$, respectively, and unchanged other parameters The
values of $r$ and $\dfrac{t_{1}}{t_{8}}$ respectively are $r$=3.62
, $\dfrac{t_{1}}{t_{8}}=2975$ for Figure 3 and $r=4.26$, $\dfrac{t_{1}}{t_{8}}=10779$
for Figure 4. Figure 4 corresponds to the case of a very weak spread
of the epidemic in the $p$-adic model.
\begin{figure}[hbt!]
\centering{}\includegraphics[scale=0.4]{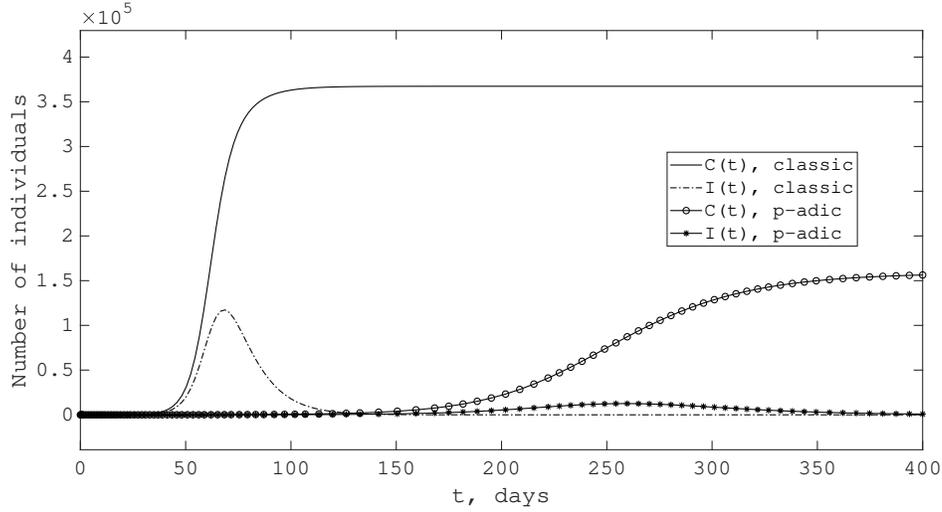}\caption{Cumulative $C\left(t\right)$ and active $I\left(t\right)$ numbers
of infected individuals in the classical and $p$-adic SIR models
for paramiters $\alpha=0.8$, $p=5$, $n=8$, $\beta=0.3$, $\gamma=0.1$. }
\end{figure}

\begin{figure}[hbt!]
\centering{}\includegraphics[scale=0.4]{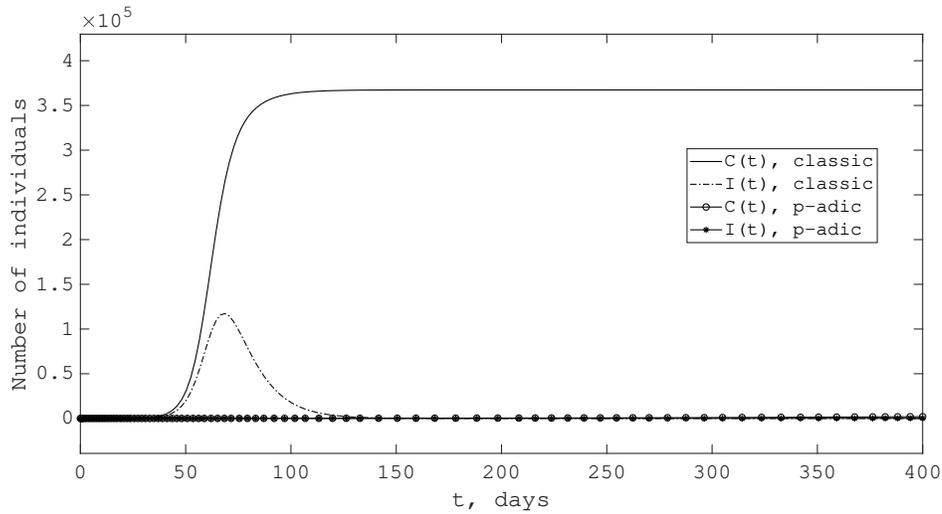}\caption{Cumulative $C\left(t\right)$ and active $I\left(t\right)$ numbers
of infected individuals in the classical and $p$-adic SIR models
for paramiters $\alpha=0.9$, $p=5$, $n=8$, $\beta=0.3$, $\gamma=0.1$. }
\end{figure}

Fig. 5 shows dependence of the cumulative number of infected individuals
$C\left(t\right)$ in $p$-adic SIR model for parameters $p=5$, $n=8$,
$\beta=0.3$, $\gamma=0.1$ and different $\alpha$. As it can be
seen from these dependencies there must exist critical $\alpha=\alpha_{0}\left(p,n,\beta,\gamma\right)$
such that at $\alpha<\alpha_{0}$ the spread of the epidemic takes
place, but at $\alpha>\alpha_{0}$ the epidemic does not spread. For
values $p=5$,$n=8$, $\beta=0.3$, $\gamma=0.1$, we have the numerical
value of $\alpha_{0}\sim1$. Unfortunately, we could not get an exact
analytical expression for function $\alpha_{0}\left(p,n,\beta,\gamma\right)$.
\begin{figure}[hbt!]
\includegraphics[scale=0.4]{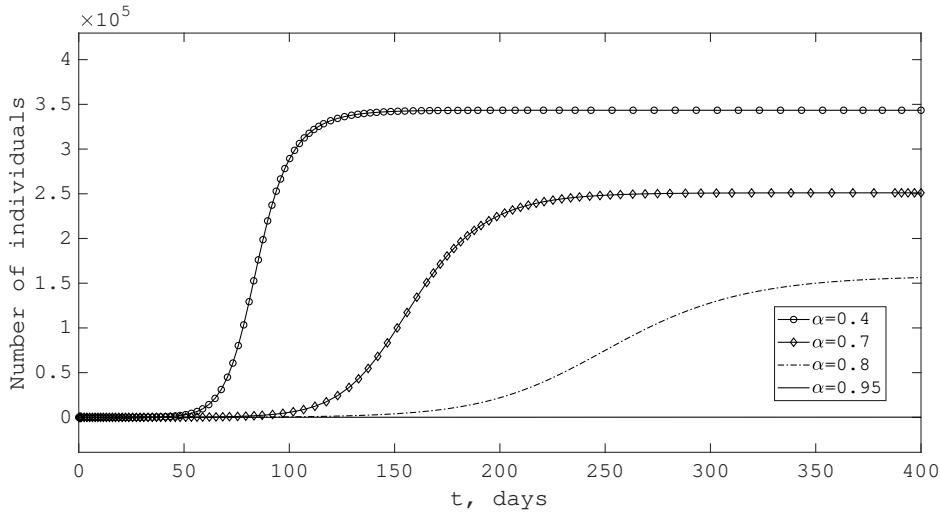}\caption{Cumulative number $C\left(t\right)$ of infected individuals in the
$p$-adic SIR model for paramiters $p=5$, $n=8$, $\beta=0.3$, $\gamma=0.1$
and different $\alpha$. }
\end{figure}

In the $p$-adic SIR model, the hierarchical isolation index (\ref{r})
can be considered as a control parameter that can be changed to control
the spread of an epidemic. To illustrate, we will consider a situation
where, as the epidemic grows, enforcement restrictions are introduced
to redistribute the time of an infectious contact between individuals
in clusters of different levels, while maintaining the average contact
time is constant. In reality, the introduction of such restrictions
means that an infectious contact between pairs of individuals with
a small ultrametric distance must be increased, and an infectious
contact between pairs of individuals with a large ultrametric distance
must be decreased. In the $p$-adic model, the introduction or strengthening
of already introduced restrictions means an increase in the $\alpha$
parameter. In Figure 6, we present dependence of the cumulative number
of infected individuals in the case of controlling their hierarchical
isolation, i.e., increasing the value of parameter $\alpha$ at time
$t_{1}$ from value $\alpha_{1}=0.5$ to value $\alpha_{2}=0.9$ with
other parameters equal to $p=5$, $n=8$, $\beta=0.3$, $\gamma=0.1$.
We see that taking an adequate enforcement restriction to increase
the hierarchical isolation index from $r_{1}=2,24$ to $r_{2}=4,26$
reduces the cumulative number of infected by more than 2 times.
\begin{figure}[hbt!]
\centering{}\includegraphics[scale=0.4]{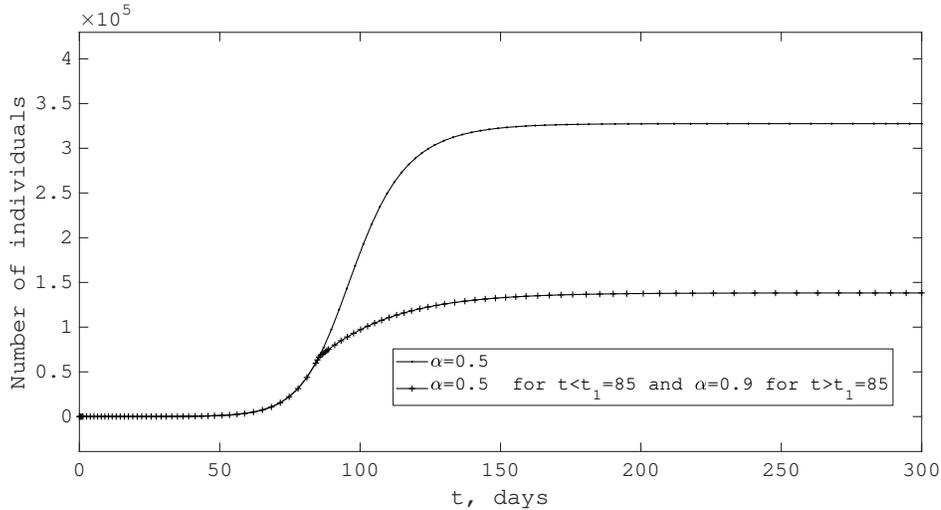}\caption{Cumulative number $C\left(t\right)$ of infected individuals in the
$p$-adic SIR model for changing the parameter $\alpha$ at time $t_{1}=85$
from $\alpha_{1}=0.5$ to $\alpha_{2}=0,9$ and for $p=5$, $n=8$,
$\beta=0.3$, $\gamma=0.1$. }
\end{figure}

\section{Conclusion }

In this paper, we have developed an ultrametric model of the epidemic
spread of infection in the population, based on the classical SIR
model. This model is also based on the concept of an infectious contact
between any two individuals in the population, which determines the
potential for transmission of infection from one individual to another.
The formalization of this concept leads to hierarchical clustering
of the population and the introduction of the ultrametric distance
on a set of individuals. The ultrametric distance between individuals
reflects the measure of transmission of infection between individuals
and is included in the ultrametric generalization equations of the
classical SIR model.

Changing the ultrametric structure of the proposed model can affect
the spread of the epidemic in the population. Therefore, in contrast
to the classic SIR model and its modifications, the proposed model
can provide recommendations for avoiding a cumulative scenario of
epidemic development based on managing the process of an infectious
interaction between clusters of individuals at various levels. This
corresponds to a certain algorithm for isolating separate social strata
of the population at the micro, meso, and macro levels.

In conclusion, we note that the ultrametric model proposed in this
paper is the basic model. It can be supplemented with various additional
scenarios, such as the possibility of re-infection, the latent period,
the time spent by individuals in the immune stage, the impact of vaccination,
etc. We reserve the implementation and application of numerous extensions
of the basic ultrametric model for future research.

\vspace{3mm}

The study was supported in part by the Ministry of Education and Science
of Russia by State assignment to educational and research institutions
under project FSSS-2020-0014.


\begin{thebibliography}{10}
\bibitem{AM} R. Anderson and R. May, Infection Deseases of Humans:
Dynamics and Control, New York, Oxford University Press, 1992.

\bibitem{DH} O. Diekmann and J.A.P. Heesterbeek, Mathematical Epidemiology
of Infectious Diseases: Model Building, Analysis and Interpretation,
John Wiley and Sons, 2000.

\bibitem{KR} M. Keeling and P. Rohani, Modeling Infectious Diseases
in Humans and Animals, Princeton University Press, 2007.

\bibitem{BS} F. Brauer F and C. Castillo-Chavez, Mathematical models
in population biology and epidemiology. Springer, 2012.

\bibitem{WB} Z. Wang, C.T. Bauch C.T., S. Bhattacharyya, et al.,
Statistical physics of vaccination, Phys. Rep., 664, 2016, 1--113.

\bibitem{KM} W.O. Kermack and A.G. McKendrick, Contribution to the
Mathematical Theory of Epidemics, Proceedings of the Royal Statistical
Society A, 115, 1927, 700--721.

\bibitem{LM} A.L. Lloyd and R.M. May, How viruses spread among computers
and people, Science, 2001292 (5520), 2001, 1316--1317.

\bibitem{MPV} Y. Moreno, R. Pastor-Satorras and A. Vespignani, Epidemic
outbreaks in complex heterogeneous networks. The European Physical
Journal B-Condensed Matter and Complex Systems, 26 (4), 2002, 521--529.

\bibitem{YWRBSWZ} R. Yang, B.H. Wang, J. Ren, W.J. Bai, Z.W. Shi,
W.X. Wang, and T. Zhou, Epidemic spreading on heterogeneous networks
with identical infectivity, Physics Letters A, 364 (3-4), 2007, 189--193.

\bibitem{Volz} E. Volz, SIR dynamics in random networks with heterogeneous
connectivity. Journal of mathematical biology, 56 (3), 2008, 293--310.

\bibitem{ALL} B. Dragovich, A. Yu. Khrennikov, S. V. Kozyrev and
I. V. Volovich, On $p$-adic mathematical physics, $p$-Adic Numbers,
Ultrametric Analysis and Applications 1 (1), 2009, 1--17.

\bibitem{VVZ} V. S. Vladimirov, I. V. Volovich and E. I. Zelenov,
$p$-Adic analysis and mathematical physics, World Sci. Publishing,
Singapore, 1994.

\bibitem{Sh} W.H. Schikhof, Ultrametric calculus. An Introduction
to $p$-adic Analysis, Cambridge Studies in Advanced Mathematics,
Cambridge University Press, Cambridge, 1984.

\bibitem{DZ} M.V. Dolgopolov, and A.P. Zubarev, Some aspects of the
$m$-adic analysis and its applications to $m$-adic stochastic processes,
$p$-Adic Numbers, Ultrametric Analysis and Applications 3 (1), 2009,
39--51.

\bibitem{SLXRHK} S. Sanche, Y.T. Lin, C. Xu, E. Romero-Severson,
N. Hengartner and R. Ke, High Contagiousness and Rapid Spread of Severe
Acute Respiratory Syndrome Coronavirus 2, Emerging infectious diseases,
26 (7), 2020.
\end{thebibliography}
\end{document}